     \documentclass[12pt]{article}
     \usepackage[margin=1in]{geometry}
     \usepackage{times}
     \usepackage{lipsum}
\usepackage{setspace}
\spacing{1.5}

\usepackage{sectsty}
\usepackage[italicdiff]{physics}

\usepackage{amsmath}

\usepackage [english]{babel}

\usepackage{etex}
\usepackage{amsthm}
\usepackage{amsmath}
\usepackage{amssymb}
\usepackage{soul}
\usepackage{calc}
\usepackage[usenames,dvipsnames]{xcolor}
\usepackage{mathptmx}
\usepackage{colortbl}
\usepackage[boxed]{algorithm2e}
\usepackage{epstopdf}
\usepackage{arydshln}
\usepackage{booktabs}
\usepackage{natbib}
\usepackage{hyperref}

\hypersetup{
	colorlinks=true,         
	linkcolor=MidnightBlue,          
	citecolor=MidnightBlue,
	urlcolor=MidnightBlue            
 }

\usepackage{breakurl}
\usepackage{bookmark}
\usepackage{etoolbox}
\usepackage{graphicx}
\theoremstyle{plain}
\theoremstyle{definition}

\usepackage[title]{appendix}
\usepackage{mathtools}
\usepackage{stackengine}
\usepackage{cancel}
\usepackage{comment}

\usepackage{authblk}
\newcommand{\be}{\begin{equation}}
\newcommand{\ee}{\end{equation}}

\title{The Hole Argument for Reference Frames}

\author[1,2]{Nicola Bamonti}
\author[3]{Henrique Gomes}

\date{}

\affil[1]{Department of Philosophy, Scuola Normale Superiore, Piazza dei Cavalieri, 7, Pisa, 56126, Italy}
\affil[2]{Department of Philosophy, University of Geneva, 5 rue de Candolle, 1211 Geneva 4, Switzerland}
\affil[3]{Oriel College, University of Oxford, OX14EW, United Kingdom}
\begin{document}

\maketitle

\begin{abstract}
\singlespacing
We exploit the results of \cite{BamontiGomes2024} concerning the dynamical (un)coupling of reference frames to gravity to analyse the role of reference frames in the Hole Argument. We introduce a new possible threat to determinism, which we call Arbitrariness Problem (\textbf{ARB}), resulting from the inherent freedom in selecting a reference frame.

\end{abstract}

\clearpage

\tableofcontents

\clearpage

\section{Observables and the Hole Argument in General Relativity}\label{sec1}

This paper explores the conceptual and technical challenges of defining local observables in General Relativity (GR), with a focus on the Hole Argument and the role of reference frames in addressing its implications. The discussion begins by investigating the notion of local observables in GR, a critical issue given the theory's diffeomorphism invariance and the difficulty of defining gauge-invariant, spatiotemporally localised quantities (Section \ref{1.1}). This leads to an examination of the Hole Argument, a problem that raises questions about the determinism of GR and physical meaning of its solutions (Section \ref{1.2}).

Building on these foundations, the paper introduces reference frames as a crucial tool for addressing challenges posed by the Hole Argument. It differentiates between coupled and uncoupled reference frames, highlighting their respective role in defining gauge-invariant observables and ensure deterministic evolution. Through this framework, in Section \ref{sec2} we introduce the Arbitrariness Problem (\textbf{ARB}) and the New Hole Argument (\textbf{NHA}), identifying novel ways in which reference frames can deepen the conceptual underpinnings of the Hole Argument.

\subsection{Introduction to Observables}\label{1.1}

Broadly, there are certain types of transformations which can be interpreted as transformations leading to redundant descriptions of physical states: these are called gauge transformations. \cite{dirac:1950,Dirac:1958a,Dirac:1964} emphasised that only gauge-invariant quantities --- those unchanged by gauge transformations --- qualify as true observables, also called \lq{}Dirac observables\rq{}. An observable of a theory is defined as a quantity that encapsulates its \lq{}physical content\rq{}, where by physical content we mean the measurable and predictable content.
Consequently, Dirac observables are associated with the results of experimental measurements of quantities predictable by the theory (either probabilistically or deterministically).

In a theory such as GR, which has as its gauge group the group of diffeomorphisms $Diff(\mathcal{M})$, it is hard to find such quantities, especially when we require them to be spatiotemporally \textit{localised}.\footnote{There are conflicting positions on taking $Diff(\mathcal{M})$ as the gauge group of GR. For example, \cite{Belot2017} argues that not all diffeomorphisms are gauge transformations. In particular, he examines the case of asymptotically flat spacetimes, where only diffeomorphisms acting trivially at the boundary (sometismes called \textit{small diffeomoprhisms}) are classifiable as gauge transformations.} The obstacle arises from the fact that the group of diffeomorphisms reshuffle the points of the manifold on which the quantities, e.g. the metric $g_{ab}$, are defined.

A well-known approach for solving the problem of defining local observables in GR is the one introduced by \cite{Rovelli_2002}, where two notions of observability are distinguished: \lq{}partial\rq{} and \lq{}complete\rq{}.
Rovelli defines partial observables as quantities \lq{}\lq{}to which we can associate a measurement procedure that leads to a number \rq{}\rq{}. Formally, they are defined as a set of gauge-dependent quantities, which coordinatise an extended configuration space.
Complete observables are obtained by relating different sets of such partial observables in a gauge-invariant way.
Importantly, complete observables are \textit{local} quantities, in a precise meaning of the term \lq{}local\rq{}.
In a nutshell, one should not think of spatiotemporal localisation in terms of points of an unobservable manifold. Instead, we adopt a relational localisation, in which fields are localised and evolve dynamically with respect to each other.

The possibility of defining local, complete observables provides a solution to the challenge of identifying local, Dirac observables in GR. In fact, complete observables naturally qualify as Dirac observables, encapsulating the theory’s  local physical content in a relationally localised manner. The reason is as follows.
In the Hamiltonian formalism, Dirac observables are quantities that commute with first-class constraints; or, alternatively, quantities that assume a single value for each set of gauge-equivalent states (the equivalence follows from the fact that Poisson brackets generate infinitesimal gauge transformations and commutation, therefore, implies gauge-invariance).
Given the (on-shell) correspondence between the Hamiltonian 3+1 symmetries (3-diffeomorphisms and refoliations \cite{Gryb2016}) and the four-dimensional spacetime diffeomorphisms of spacetime (\cite{Lee1990}), we have, in the case of GR, a neat correspondence between Dirac and complete observables (\cite{Dittrich2006,Dittrich2007}).
In common terminology, we call \textit{relational observables} the complete observables where one of the two partial observables plays the role of reference frame for the other.
From this point of view, one can think of the complete and partial observables programme as allowing us to \lq{}de-parameterize \rq{} evolution purely in terms of dynamically coupled reference frames (\cite{Brown1995,Thiemann-k-essence, Tambornino2012,Bamonti2024}).

An intriguing result in the foundations of the concept of observability is found in \cite{BamontiGomes2024}.
Invariance under diffeomorphisms which they name \textbf{(RI)} (from reshuffling-invariance), characterises relational quantities. Their analysis shows that \textbf{(RI)} and the property of gauge-invariance, which they name \textbf{(GI)}, are intricately connected, but distinct.
In particular, \textbf{(GI)} imposes an additional requirement: deterministic evolution.
Thus, although \textbf{(RI)} is a necessary condition for a complete observable, it is not sufficient to ensure \textbf{(GI)}, which is also necessary to define a complete observable. A quantity achieves \textbf{(GI)} status only when \textit{both} \textbf{(RI) }and deterministic dynamics \textbf{(DET)} are  satisfied.
The authors summarise such implications with the following formula (ivi, p.18):
$$\textbf{(GI)} \leftrightarrow [\textbf{(RI)}\land \textbf{(DET)}].$$
This apparatus of definitions makes it possible to distinguish between \textit{relational observables} (i.e. \textbf{(RI) }observables) and \textbf{(GI)}, relational observables.

Furthermore, the authors argue that for \textit{partial observables} to be properly defined within a physical theory, they must exhibit relational behaviour, implying that they must be physically instantiated, and so must serve as a relational anchor. Furthermore,  being associated with a measurement procedure, they must satisfy \textbf{(RI)}. However, this is not enough to qualify as a \textit{bona-fide} partial observable. In fact, partial observables must be dynamically coupled to each other. Only in this way can their relationship constitute a \textit{bona-fide} \textbf{(GI)}, \textit{complete observable}.

\subsection{The Hole Argument}\label{1.2}

The difficulty in defining local observables in GR is closely linked to the problem of determinism, as illustrated by the (in)famous hole argument.
It is broadly agreed that one of the earliest conceptual problems in the foundations of GR took the form of what is now known as \lq{}the hole argument\rq{} (\cite{Earman1987-EARWPS}).
One way of posing the question raised by the argument is: how should we interpret the differences between the metric $g_{ab}$ and a symmetry-related one $[d^* g]_{ab}$, sharing the same initial data? In particular: do the two metrics represent two distinct physical possibilities?
These questions have implications for the physical determinism of the theory and the notion of measurability, which we will now explore in more detail, following the presentation of \cite{HOLE}.

\paragraph{The Problem of Indeterminism.}\label{sec1.1}
Given a solution $g_{ab}$ (a metric) of GR over the manifold $\mathcal{M}$, with initial data $\Delta^g$ on a Cauchy surface $\Sigma \subset \mathcal{M}$ (a global instant), we can obtain infinitely many other solutions, with the same $\Delta^g$, threatening indeterminism.  The alternative solutions are obtained by smoothly 'reshuffling' the manifold points with $d: \mathcal{M} \rightarrow \mathcal{M}$, any one-to-one map on spacetime with $d$ and $d^{-1}$ smooth. In more detail, since  $d$ takes smooth curves to smooth curves, it will induce a map on tangent vectors and co-vectors, denoted by $d_*$ and $d^*$ respectively, and also on  their tensor products; so it induces such a map on the metric tensor. If we pick a $d$ such that  $d^*_{|_\Sigma} =\mathsf{Id}$, we 'slide' the metric's profile of values, obtaining $[d^* g]_{ab}$, and leaving the values of $g_{ab}$ at $\Sigma$ untouched, and equal to $\Delta$. Both the original metric and the metric with the profile of values slid by the smooth reshuffling are solutions of the Einstein equations, since these equations covary under this reshuffling. So these are both dynamical and spacetime symmetries.   We will call this problem of indeterminism the \lq{}\textit{standard hole argument}\rq{} (\textbf{SHA}).

\paragraph{The Underdetermination Problem}\label{subsec1.2}
Following \cite{HOLE}, the \lq{}\textit{Underdetermination Problem}\rq{} is related to, but distinct from the \textbf{SHA}.

To expand on this difference, the authors argue that empirical data is relational in character: 
\begin{quote}
    In general, explicit representation, at least in an idealised manner, of the observer, as a physical system within spacetime, naturally leads to what one might call an \lq{}immanent\rq{} conception of empirical (in)equivalence: \lq{}Two models are empirically distinct just in case there are relevant relational differences between the field configurations in each.\rq{} \cite[p.10]{HOLE}
\end{quote} 
Then, underdetermination problem ensues when empirical data, being relational, does not allow observers to tell whether they live in $\langle\mathcal{M},g_{ab}\rangle$ or in the relationally identical $\langle\mathcal{M},[d^*g]_{ab}\rangle$.

\subsubsection{Three solutions to the SHA}

\paragraph{The Relational Camp.}

Einstein struggled with the conundrum of the classical hole argument in the final years before the birth of GR, coming finally to conclude that \lq\lq{}physical significance should only be attributed to point coincidences\rq{}\rq{} (see \cite{Stachel1989-STAESF,Giovanelli2021} and references therein), a strategy that has a relational character.
Assuming that only relational quantities can have physical significance, the difference between $g_{ab}$ and its symmetry-related model $[d^* g]_{ab}$ becomes merely mathematical,  not physical. In other words, it is a gauge difference.

In this work, we are inspired by this strategy introduced by Einstein to address the hole argument and we argue that the use of reference frames formalism this intuitive response, and forecloses the hole argument.
The role of reference frames in GR has an extensive philosophical literature (see \cite{Bamonti2023} and references therein).

At this early stage, it is sufficient to define a reference frame as a physical system whose fixed set of relations provide a local diffeomorphism $U \rightarrow \mathbb{R}^4$, for some $U \in \mathcal{M}$, which \textit{uniquely} assigns four numbers to each point in $U$.
In this way, a tensor defined on the manifold, such as the metric $g_{ab}$, can locally be parametrised by the chosen reference frame.

\begin{figure}[h!]
    \centering
    \includegraphics[scale=0.22]{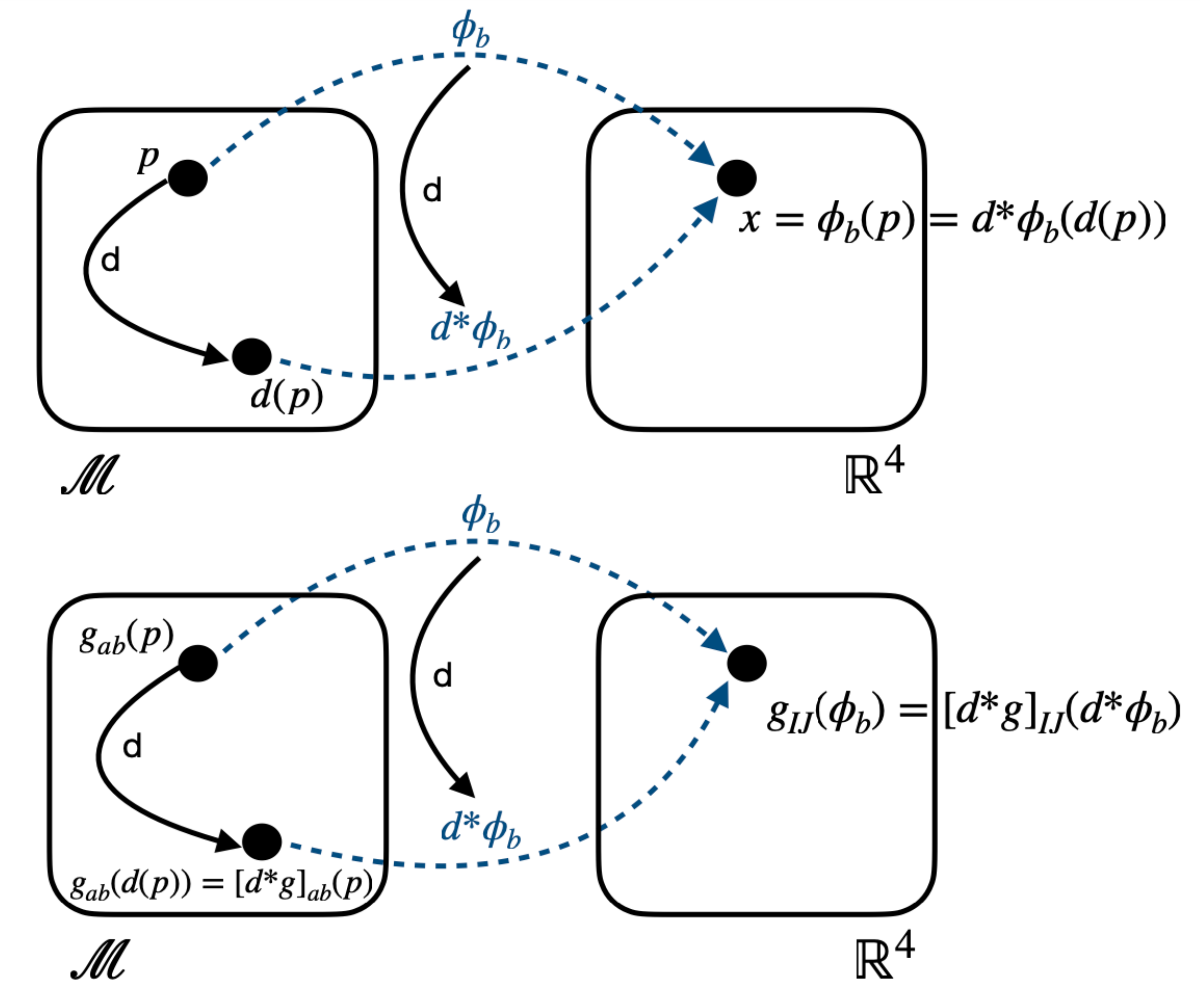}
    \caption{Action of diffeomorphism $d$. Note that $d$ acts on \textit{both} the coupled fields $(\phi_b,g_{ab})$.}
    \label{figdvsd}
\end{figure}

For example, using a set of  linearly independent four scalar fields $\{\phi^{(I)}\}_{I=1,\dots,4}$ , satisfying Klein-Gordon equations, on some $U \subseteq \mathcal{M}$, we can define $g_{IJ}(\phi):= \big[(\phi^{(I)})^{-1}\big]^* g_{ab}$, where $\big[ \bullet \big]^*$  denotes the pullback and  $g_{IJ}(\phi)$ the components in the frame $\{\phi^{(I)}\}$ of the abstract metric tensor $g_{ab}$.
The quantity $g_{IJ}(\phi)$ is a local, \textbf{(GI)} \textit{relational observable}.  To demonstrate that $g_{IJ}(\phi)$ always satisfies \textbf{(RI)}, we present the following proof\footnote{We use the standard relations: $(f \circ h)^{-1}=h^{-1} \circ f^{-1}$ and $(f \circ h)^*=h^*\circ f^*$.} (see also Figure \ref{figdvsd} which provides a graphic example of the \textit{diagonal} action of the diffeomorphism $d$):
\begin{proof}
\begin{align*}
\big[(d^*\phi^{(I)})^{-1}\big]^* [d^*g]_{ab}&=\big[(\phi^{(I)} \circ d)^{-1}\big]^* [d^*g]_{ab}
\\&= \big[d^{-1} \circ (\phi^{(I)})^{-1}\big]^* [d^*g]_{ab}
\\&= \big[(\phi^{(I)})^{-1}\big]^* \circ \big[d^{-1}\big]^*[d]^*g_{ab} =: g_{IJ}(\phi).
\end{align*}
\end{proof}

In this paper, we make use of the classification of reference frames in GR found in \cite{Bamonti2023} and later refined in \cite{BamontiGomes2024}.
Within GR, the authors distinguish between reference frames dynamically coupled with gravity (coupled reference frames: \textbf{CRFs}) and reference frames uncoupled with gravity (uncoupled reference frames: \textbf{URFs}).
In particular, it is only the former class that allows the definition of \textbf{(GI)}, relational observables. In the case of \textbf{URFs}, the quantity $g_{IJ}(\phi)$ is a relational (i.e. \textbf{(RI)}) observable, but not a \textbf{(GI)}, relational observable.

It is easy to show that only in the case in which the set of $\{\phi^{I}\}$ constitues a \textbf{CRF}, determinism is guaranteed, thus foreclosing the hole argument.\footnote{We will return to this point in the next section, highlighting how the underdetermination problem is also solved by the use of \textbf{CRFs}.}
In fact, notice that, for \textbf{CRFs}, when the pair $(g_{ab},\phi^{(I)})$ is a possible solution, then $([d^*g]_{ab},\phi^{(I)})$ is not, for a generic diffeomorphism $d \in Diff(\mathcal{M})$.
Thus, a choice of $\phi^{(I)}$ (rather than any of its isomorphic distributions) and initial data give us a \textit{unique} representation for $g_{IJ}(\phi)$.
For this reason, \textit{at most one} of all of the isomorphic copies of $g_{ab}$ is compatible with each $\phi^{(I)}$ in its isomorphic class. Conversely, for \textbf{URFs}, both the combinations $([d^*g]_{ab},\phi^{(I)})$ and $(g_{ab},d^*\phi^{(I)})$ are possible solutions.\footnote{See \cite{BamontiGomes2024} for an in-depth study of the dynamical symmetry group for such uncoupled fields. In particular, the authors show that $([d^*g]_{ab},\phi^{(I)})$ and $(g_{ab},d^*\phi^{(I)})$ are possible solutions $\forall d \in Diff(\mathcal{M})\times Diff(\mathcal{M})$. Therefore, the group of dynamical symmetries must be expanded with respect to the case of coupled fields, where the dynamical symmetries are  $d \in Diff(\mathcal{M})$.}

\paragraph{The Substantivalist Camp.}

Other possible solutions to the hole argument have emerged in modern literature, some of them aimed at rescuing some of the intuitions behind substantivalism: the metaphysical doctrine according to which, very broadly, space-time exists as a \textit{sui generis} kind of \lq{}substance\rq{}, independent of material content (see \cite{Brown2013} for an analysis on the category of \lq{}substance\rq{}).
However, the hole argument poses a serious challenge to substantivalism, particularly to the naive or \textit{haecceitistic} version, which attributes a primitive identity (\textit{haecceitas}) to spacetime points (\cite{Earman1987-EARWPS}).
Nonetheless, there are compelling motivations to rescue substantivalism. First, substantivalism preserves the explanatory power of spacetime as a \textit{sui generis} kind of \lq{}substance\rq{} capable of influencing physical processes, as evidenced in GR;\footnote{See \cite{Brown2005-kq}'s dynamical view on spacetime for an alternative stance on the role of spacetime, especially in special relativistic theories.} second, it provides a robust framework for understanding inertial effects.

Building on these strengths, several refined substantivalist frameworks have been proposed to resolve the challenges posed by the Hole Argument while preserving the advantages of substantivalist ontology. These approaches aim to reconcile the explanatory power of spacetime, avoiding the pitfalls of naive substantivalism.

One prominent example is ‘\textit{sophisticated} substantivalism,’ which reinterprets spacetime points in a way that avoids the indeterminism implied by naive substantivalism while retaining their ontological significance (\cite{HOLE}).
This is a particular example of a broad position on symmetries called \textit{sophistication}.
Sophistication takes isomorphism to provide a standard of physical equivalence between models, but does not simultaneously seek an elimination of symmetry-related models from the formalism. 
Sophisticated substantivalism upholds the intuitive (substantivalist) picture of spacetime as a manifold, endowed with geometry and various matter fields, but this is done within an \textit{anti-haecceitist} construal of spacetime points in which they have no \lq{}primitive identity\rq{}. According to this construal, spacetime point can \textit{only} be individuated through the mesh of properties and relations in which they stand with other points.
Within \textit{sophisticated substantivalism}, it is important to note that \lq{}\lq{} the denial of primitive identity for spacetime points is compatible with the existence of a ‘bare’ spacetime manifold composed of ‘numerically’ distinct spacetime points. That is, the numerical distinctness of spacetime points can be stated without reference to the points’ ‘primitive identity’ (i.e. without the use of free, singular terms denoting individual elements of the set); indeed,  distinct points can have all the same qualitative properties--—so the view is not committed to the principle of identity of indiscernibles in this sense.\rq{}\rq{} \cite{GomesNoether}.

A similar-minded view on symmetry-related models of general relativity, inspired by \cite{Weatherall2018-WEARTH-2}, has been called the \lq{}drag-along\rq{} proposal, as described by \cite{GomesHole1,GomesHole2}.
As \cite{Weatherall2018-WEARTH-2} writes: 
\begin{quote}
    When we say that $\langle\mathcal{M},g_{ab}\rangle$ and $\langle\mathcal{M},\hat{g}_{ab}\rangle$ are isometric spacetimes, and thus that they have all of the same invariant, observable structure, we are comparing them relative to [the isometry \dots]. If one only considers [the isometry], no disagreement arises regarding the value of the metric at any given point, since for any point $x\in \mathcal{M}, g_{ab}(x) = \hat{g}_{ab}(f(x))$ by construction (ivi, p. 336).
    \end{quote}
More generally, Gomes claims that in order to compare what two isomorphic metrics $g_{ab}$ and $[d^*g]_{ab}$ say about points $p\in \mathcal{M}$, in principle we could use any diffeomorphism that \lq{}reshuffles points\rq{}. However, the only comparison that would preserve physical content \textit{pointwise}, and not only for the entire model, is that which gives rise to the isometry, namely $d$. Thus, given $d\in Diff(M)$, $g_{ab}$ and $[d^*g]_{ab}$, for every $p\in M$, we should compare $g_{ab}(p)$ with $[d^*g]_{ab}(d(p))$. And indeed, the two tensors are numerically equal and represent the same physical situation point by point. The drag-along proposal is in line with the anti-haecceitist doctrine since it disallows non-qualitative identification of points across isomorphic models.
That, is, we should not compare models using both the identity on (points of) $\mathcal{M}$ and the (pull-back of the) diffeomorphism on the tensor fields $g_{ab}$.
This is one way to avoid what \cite{HOLE} call the \lq{}equivocation argument\rq{}, upon which the hole argument rests: an illegitimate equivocation between two maps, the isometry and the identity.

We will show in $\S$ \ref{sec2.1} how the drag along proposal can be reconciled with our strategy to address the classical hole argument, based on reference frames formalism.

\section[The Hole is shut. It was made by those who are Dead, and the Dead keep it. The Hole is shut]{The Hole is shut. It was made by those who are Dead, and the Dead keep it. The Hole is shut\footnote{This should not be taken literally. It is a rephrasing of the famous quote: \lq{}\lq{}\textit{The way is shut. It was made by those who are Dead, and the Dead keep it, until the time comes. The way is shut.}\rq{}\rq{} from J.R.R. Tolkien, The Return of the King (1955). In reality, many thinkers who have \lq{}made\rq{} the hole argument are alive and well.}} \label{sec2}

\subsection{The case of \textbf{CRFs}}\label{sec2.1}
As argued in the previous section, \textbf{CRFs}  solve the problem of defining local, \textbf{(GI)} complete observables, by rendering the dynamics deterministic (Section \ref{sec1}). Thus,  they avert the threat of indeterminism plaguing the \lq{}standard\rq{} hole argument (\textbf{SHA}).

Nonetheless,  the freedom of the choice of which physical system will play the role of the reference frame remains.
In general, such a choice of a physical field as a reference frame is only local. That is, we cannot find a single suitable dynamically coupled reference frame that covers the whole $\mathcal{M}$.
Because it stands out for its straightforward visualisability, an example of a \textbf{CRF} that we will use to analyse the role of reference frames in the hole argument is represented by the set of the so-called \textit{GPS coordinates}, introduced in \cite{RovelliGPS}\footnote{The idea is to consider the system formed by GR coupled with four test bodies, referred to as satellites, which are deemed point particles following timelike geodesics, meeting at some (starting) point O. Each particle is associated with its own proper time $\phi$. Using light signals from the satellites, we can uniquely associate four numbers $\phi^{(I)}, I=1,2,3,4$ to each spacetime point $P$ in the appropriate region. These four numbers represent the four physical variables that constitute the \textbf{CRF}. Physically they constitute the lightlike distance between the intersection points with the past lightcone of $P$ and the starting point $O$.}.
In the following, we take two distinct sets of GPS reference frames, which allow us to define points in the same local region of spacetime relationally, but in two different ways.
We define the \lq{}red\rq{} set of GPS satellites, $\{r^{(I)}\}$, and the \lq{}blue\rq{} one $\{b^{(I)}\}$. With these, we construct the \lq{}red\rq{} reference frame $\{\phi_r^{(I)}\}$ and the \lq{}blue\rq{} one $\{\phi_b^{(I)}\}$, imagining, only in order to aid visualisation, that they cast \lq{}blue\rq{} and \lq{}red\rq{} \lq{}physical parametrisations\rq{} on that region of spacetime. It is important to emphasise that the reference frames are obtained from  two \textit{physically distinct} dynamical systems, or sets of physical objects: e.g. two different sets of GPS satellites.\footnote{Our discussion remains valid if the eight scalar fields are considered to represent two \textit{distinct} sets of real Klein-Gordon scalar fields, each satisfying $\Box \phi=0$. We choose to use GPS frames because they represent a realistic example of a reference frame.}
What we mean is that the two sets of GPS frames $(\phi_r^{(I)},\phi_b^{(I)})$ satisfy two \textit{different sets of initial conditions}, for a \textit{given} metric $g_{ab}$.\footnote{NB: this does not mean that the metric is \textit{given} in the sense that it is an \textit{absolute field}: i.e. the same (up to isomorphism) in every DPM, or a \textit{fixed field}: i.e. the same in every KPM (\cite{Anderson1967-en}, \cite{James_Read2023-mk}). It only means that we do not consider back-reaction. We must not have a \textit{given} metric in the above meanings, because that would reduce the discussion to a theory in a curved background. Here, we deal instead with GR which is a background independent dynamical theory of the gravitational field.} {The model $\langle g_{ab},\phi_r^{(I)},\phi_b^{(I)}\rangle$ is accompained by the set of initial data $(\Delta^g,\Delta^{\phi_r}_g,\Delta^{\phi_b}_g)$ for $g_{ab}$, $\phi_r^{(I)}$ and $\phi_b^{(I)}$, respectively.}\footnote{Note that in case of what \cite{Bamonti2023} call \textbf{RRFs}, that is backreacting reference frames, the initial data would be: $(\Delta^g_{\phi_r,\phi_b},\Delta^{\phi_r}_{g,\phi_b},\Delta^{\phi_b}_{g,\phi_r})$. The initial conditions of the metric and material fields cannot be chosen independently of each other, since they are intertwined through the EFEs.}
Moreover, \textit{we are considering diffeomorphisms that do not change the initial data}. This restriction implies no loss of generality.\footnote{To see this, consider a particular diffeomorphism $d$, whose pull-back takes initial data $\Delta \rightarrow \Delta'$. Suppose this new initial data provides a solution to the KG equation with respect to the original metric $g_{ab}$ (which recall we have left untouched). But that solution will be what it will be:  generically it won't be $d^*\phi$ throughout spacetime. And even supposing that, by coincidence, $d^*\phi$ is the solution everywhere, for each such $d$, there is a non-denumerable set of diffeomorphisms that match $d$ on the initial data and that differ elsewhere. And it is easy to see that, generically, these can't also produce a solution to the KG equation: since $(g_{ab}, d^*\phi)\simeq ([d^*g]_{ab}, \phi)$, and the KG equation using $g_{ab}$ and $[d^*g]_{ab}$ will have different terms (e.g. different Christoffel symbols) appearing for an equation for the same solution, $\phi$, so generically can't both be satisfied. A quick way to see that this restriction is unimportant for questions of (in)determinism, note that by joining the set of diffeomorphisms that are non-trivial at a Cauchy hypersurface $\Sigma$ but \textit{time-independent }(in some coordinates) with the set that preserves $\Sigma$ but are otherwise arbitrary, we recover the full set of diffeomorphisms. But time-independent gauge transformations do not give rise to indeterminism (see \cite{Wallace_2003}), and so we have captured the only set that is relevant for this question: that of $\Sigma$-preserving, but arbitrary diffeomorphisms.}

But now, we are left with another redundancy with respect to the gauge redundancy of the \textbf{SHA}, the \lq{}\textit{Arbitrariness Problem}\rq{} (\textbf{ARB}):
\begin{description}
\item[\textbf{ARB:}] Given a metric field $g_{ab}$,  in order to write \textit{different} local Dirac-observable metrics, in principle we have total freedom to choose between the red or the blue reference frame, obtaining $g_{IJ}(\phi_r)$ and  $g_{IJ}(\phi_b)$. Thus, the resolution of the \textbf{SHA} still leaves open a possible worry about the arbitrary choice of reference frames: the blue or the red set.
\end{description}

\begin{description}
    \item[\textbf{ARB resolution:}] Having assumed that they overlap on the entire spacetime manifold $\mathcal{M}$ (a clearly unrealistic supposition),\footnote{Since we need multiple fields to cover the whole spacetime manifold, a compatibility condition must be imposed on their overlap. \label{fn25}} one can define a map relating the two frames which is essentially the same as a change of coordinates. Let us rewrite the red set and the blue set in the more familiar notation for coordinates: $\phi_r^{(I)}:=X^I_r$ and $\phi_b^{(I)}:=X^{I'}_b$. The two local \textbf{(GI)} observables $g_{IJ}(\phi_r):= g_{IJ}(X^I_r)$ and $g_{IJ}(\phi_b):=g_{I'J'}(X^{I'}_b)$ are related by a map \textbf{m} which acts as follows:

\begin{equation}
   \text{\bf{m}}:g_{IJ}(X^I_r) \rightarrow g_{I'J'}(X^{I'}_b)=\cfrac{\partial X^I_r}{\partial X^{I'}_b}\cfrac{\partial X^J_r}{\partial X^{J'}_b} g_{IJ}(X^I_r) \label{eq1}
\end{equation}

It is evident that this is a mere passive diffeomorphism transformation.\footnote{We believe this neatly addresses the criticism that the choice of a reference frame, as it is related to a gauge-fixing procedure \citep{Bamonti2023}, may constitute a break of the [gauge] covariance of the theory.
This is clearly not the case.
This answer seems clearer to us than the one given by \cite[p.15]{GomesNoether}, who follows \cite{Komar}, according to whom \lq{}\lq{}Though we found this dressed observable by employing a gauge-fixing, we need not think of it in those terms: a dressed quantity [\dots] is just invariant under diffeomorphisms, full stop\rq{}\rq{}. Another issue is whether the choice of a reference frame spoils the gauge invariance of the theory by introducing gauge dependence, namely a dependence on the reference frame chosen (see \cite{Wallace2024}). In our opinion, the choice of a reference frame actually makes physics frame-dependent, but this is not a major concern. A frame-dependent description is still gauge-invariant. This is a subtle point and is based on the crucial distinction between d and \textbf{d} maps (see below). \label{fninvariance}}

Given the 1-1 correspondence between active and passive diffeomorphisms, from the active point of view, the relation between $g_{IJ}(\phi_r)$ and $g_{IJ}(\phi_b)$ can be understood as follows:  suppose that $\phi_b$ takes a point $p$ to the 4-tuple $(a,b,c,d)$ defining a point $x \in \mathbb{R}^4$. And $\phi_r$ takes $p$ to the point $y=(a’,b’,c',d')$. We could also see the point $x=(a,b,c,d)$, which refers originally to values in $\phi_b$, as referring to values in $\phi_r$. But these values will not be mapped back to $p$ according to $\phi_r^{-1}$: they are mapped to a different point, $p’=q=\phi_r^{-1}(x)$. If one does this with every point in the relevant domain of $\mathbb{R}^4$, one gets an active diffeomorphism $\textbf{d}:=\phi_r^{-1}\circ\textbf{m}\circ\phi_r$ that maps between points of $\mathcal{M}$, where \textbf{m} is the passive map from $\phi_r$ to $\phi_b$.\footnote{One could also see the active diffeomorphism as a \textit{right action} $\phi_r:=\phi_b \circ \textbf{d}^{-1}$, whereas the passive diffeomorphism as a \textit{left action} $\phi_r:=\textbf{m}^{-1} \circ \phi_b$. \label{fnleftright}} Consequently, one recovers the active diffeomorphism relating $g_{IJ}(\phi_r)$ and $g_{IJ}(\phi_b)$. \footnote{We can rehearse the usual mathematical argument to introduce the 1-1 correspondence between active and passive diffeomorphisms: for a given $U\subset \mathcal{M}$ as the domain of a chart $\gamma: U\rightarrow \mathbb R^k$, and  given an active diffeomorphism $d\in\text{Diff}(U)$, we can find a passive diffeomorphism 
\begin{equation} m:=\gamma\circ d\circ \gamma^{-1} \in\text{Diff}(\mathbb R^k).\end{equation}
And, conversely, given $ m\in \text{Diff}(\mathbb R^k)$, we can find an active diffeomorphism
  \begin{equation}\label{eq:passive_active} d:=\gamma^{-1}\circ m \circ \gamma \in\text{Diff}(U).\end{equation}
This active diffeomorphism should be interpreted as follows: given coordinates $x$ for a point $p$, i.e. $\gamma(p)=x$, we can, under a different chart, map $x$ to a different point, $q$, such that $\gamma\rq{}(q)=x$. Different coordinate charts will ascribe a different value, a different spacetime point, to the same coordinates; so, having chosen a chart,  a change of coordinates gives rise to a unique active diffeomorphism acting on the spacetime points. 
Agreed, we could have $d(U)\cap U=\emptyset$, in which case we clearly would not be able to register this diffeomorphism using solely a coordinate transformation in the domain of the single chart $\gamma$ (see \cite[fn.7]{Norton1989}).
Having said that, if we restrict the diffeomorphisms to be connected to the identity, we need to only consider their generators, which are the infinitesimal flow of vector fields. And though these vector fields may be non-trivial at the boundary of the chart, they can still be represented within the charts, and thus at their intersection as well. Thus, given any atlas for the manifold---any covering of the manifold by a finite number of charts---we can patch together any infinitesimal active diffeomorphism using the infinitesimal passive diffeomorphisms in each chart of the atlas, and, by integration, recover the 1-1 correspondence between the active and the passive viewpoint.\label{fn30}}
The observables $g_{IJ}(\phi_r)$ and $g_{IJ}(\phi_b)$ say different things.
But it does not matter whether we choose the blue or the red set to define \textbf{(GI)} observables, as we have a map to translate one choice into the other.\footnote{A different procedure for describing a change of reference frame can be found in \cite{QuantumHole}.
Here, the authors use the abstract formalism of the fibre-bundle and the notions of \textit{representational convention}, understood as a \textit{section} that \lq{}cuts\rq{} the fibres of the isometry group once and only once, and \textit{counterpart relation} between models (\cite{GomesReprConventions}).}
\end{description}

Despite this, it is important to point out that it is not straightforward to interpret this change of reference frame as a change in the way we choose to represent the \textit{same} physical state. In this case, a change of chart is \textit{not} a mere notational change: a change of representation deprived of any ontological significance.\footnote{This is in contrast to the \lq{}perspectivalist\rq{} position advocated in \cite{Giacomini2019} and \cite{QuantumHole} (also applied to the quantum regime), according to which the covariance of physical laws under a change of reference frame does not affect the physical situation. Or stating otherwise: a change of reference frame does not change the physical content of our description.}
Each of the observables $g_{IJ}(\phi_r)$ and $g_{IJ}(\phi_b)$ represents a \textit{fully-fledged} physical situation in terms of different quantities that at a surface level might have little resemblance to each other. 
What we have shown in eq.(\ref{eq1}) is only that the (representations of the) two physical situations are connected by a diffeomorphism. In other words, even if the two \textbf{(GI)} local observables represent two at first glance independent physical situations, we know which map to use to translate one into the other. But there may be no \lq{}objective, view-from-nowhere\rq{}, from which these two perspectives derive: in short, the point is that, each different set of fixed initial conditions for $\phi_r^{(I)}$ and $\phi_b^{(I)}$ are \textit{univocally} associated each to a different physical solution, and these solutions, for the region which they map diffeomorphically to $\mathbb{R}^4$, are related by the diffeomorphism \eqref{eq1}. However, thankfully, we are not condemned to have several possible solutions from \textit{the same} initial data.\footnote{While \textbf{CRFs} do not leave open any indeterminism problem, or any arbitrariness problem, they might leave the Underdetermination Problem open: having chosen a reference frame, different solutions are related by a diffeomorphism \textbf{d}. Can the empirical relational data distinguish between these diffeomorphism-related solutions?
But that possibility is easily dismissed:
In fact, the underdetermination problem of \cite{HOLE} is based on two non-relational quantities, which do not satisfy \textbf{(RI)}: $g_{ab}$ and $[d^*g]_{ab}$.
In our case $g_{IJ}(\phi_r)$ and $g_{IJ}(\phi_b)$ are relational (thus, \textbf{(RI)}) , even \textbf{(GI)}, quantities. Consequently, empirical data \textit{can} distinguish between the two physical solutions. That is,  if we can take the choice of blue or red sets of GPS to be univocal, and to be unambiguously settled by the practitioner---like many other settings in the context of application of a theory---then there is no underdetermination left.
Of course, \textbf{ARB} and this version of underdetermination problem are intimately linked, in the sense that the former is \textit{solved} from the presence of \textbf{d}, while the latter \textit{arises} from the presence of \textbf{d}.}

Finally, notice that each reference frame is available in a restricted subset of the configuration space of GR, and each \textit{defines} a notion of locality with respect to itself (namely, via the map to $\mathbb{R}^4$).

\paragraph{External diffeomorphism.}

We have indicated with \textbf{d} in bold the specific active counterpart of the passive \textbf{m}-map relating $g_{IJ}(\phi_r)$ and $g_{IJ}(\phi_b)$. On the other hand, by $d$ not in bold, we refer to the usual action of an active diffeomorphism on the model $\langle g_{ab},\phi_r^{(I)}, \phi_b^{(I)} \rangle$.
The reason why it is useful to stress this distinction explicitly is that one could worry that since $g_{IJ}(\phi_b)$ is \textbf{(GI)} it is also \textbf{(RI)}, so in principle the action of \textit{any} diffeomorphism $d \in Diff(\mathcal{M})$ should leave it unchanged, however, the active diffeomorphism \textbf{d} acts as: $Diff(\mathcal{M}) \ni \textbf{d}: g_{IJ}(\phi_b) \to g_{IJ}(\phi_r)$. Here's a short proof:
\begin{proof}
Given $g_{IJ}(\phi_b):= \big[ \phi_b^{-1} \big]^*g_{ab}$ and $\phi_r^{-1}:= \textbf{d} \circ \phi_b^{-1}$ (see footnote \ref{fnleftright}), and rewriting the action of \textbf{m} as the action of \textbf{d}, that is: \textbf{m}$^{-1}:=\big[ \phi_b\circ \textbf{d} \circ \phi_b^{-1} \big]$, we have:
\begin{align*}
(\textbf{m}^{-1})^*[g_{IJ}(\phi_b)]&=(\textbf{m}^{-1})^*\big[ (\phi_b^{-1})^*g_{ab}\big]\\
&= \big[ \phi_b^{-1}\circ \textbf{m}^{-1}\big]^*g_{ab}\\
&= \big[\phi_b^{-1 }\circ \phi_b\circ \textbf{d} \circ \phi_b^{-1}\big]^*g_{ab}\\
&=\big [ \phi_r^{-1}\big]^*g_{ab}\\
&= g_{IJ}(\phi_r).
\end{align*}
In the second line of the proof we have used a fundamental property of pullback, which states that the pullback of a pullback corresponds to the pullback with respect to the composition of the maps. 
\end{proof}
How does a diffeomorphism change and not change the same quantity?
We have to be careful here. The distinction between d and \textbf{d}, is crucial.

\begin{figure}[h!]
    \centering
    \includegraphics[scale=0.22]{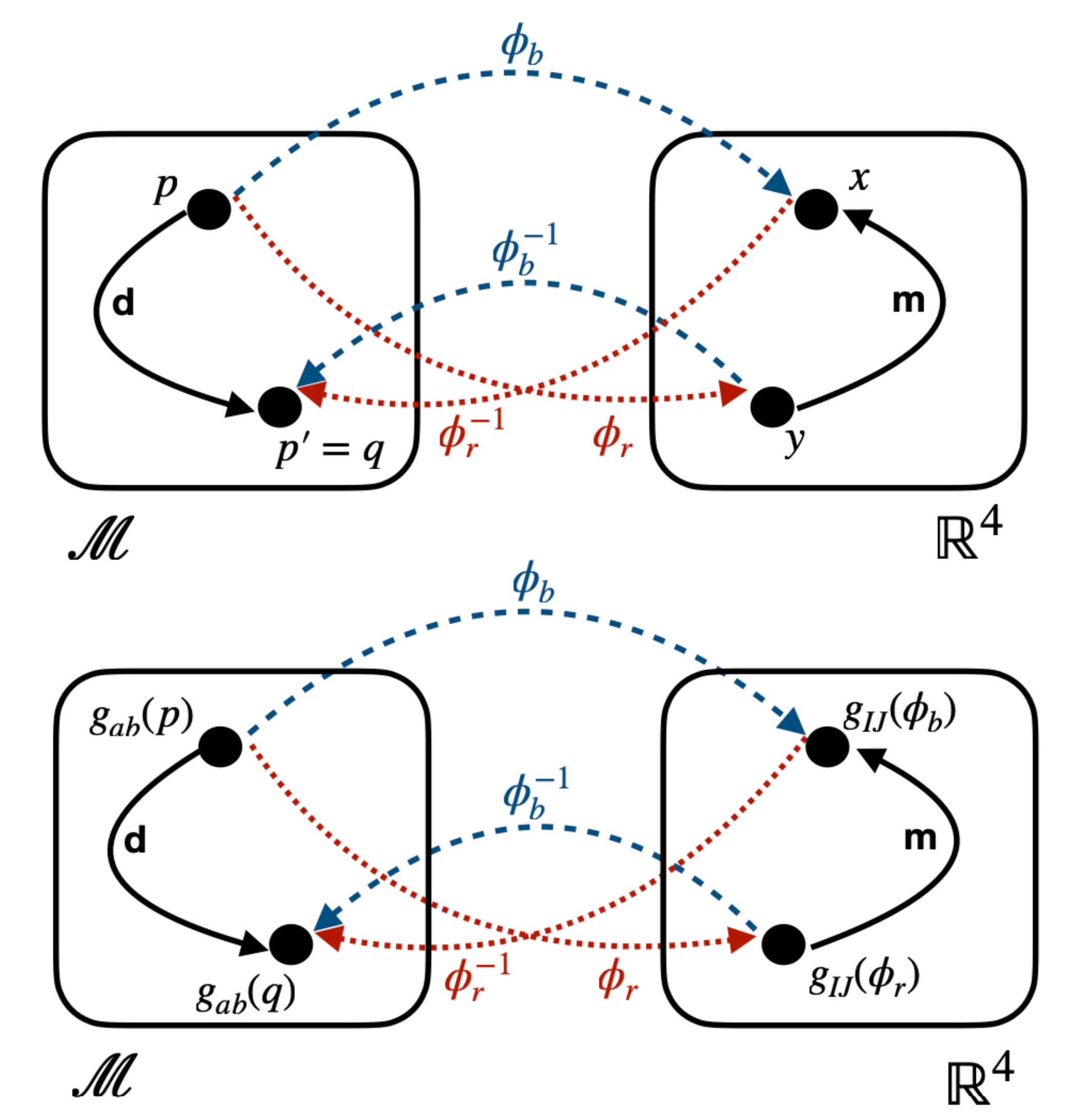}
    \caption{1-1 correspondence between the passive map \textbf{m} and the active map \textbf{d}.}
    \label{fig1}
\end{figure}

Take $g_{IJ}(\phi_b)$. It is a \textbf{(GI)} quantity, understood as functional of the metric and the GPS fields: $\mathcal{F}(g,\phi_b)$.
However, this is not to say that we couldn’t find a \textit{different} \textbf{(GI)} functional $\mathcal{F}(g,\phi_r)$ related to $\mathcal{F}(g,\phi_b)$ by a map like \textbf{d}, which we call an \textit{external diffeomorphism}.
The transformation \textbf{d} acts \textit{directly} on the already constructed \textbf{(GI)} observables, changing frames and getting us to a different and \textit{new} \textbf{(GI)} observable. Indeed, this \textbf{d} relates points with different relational properties e.g. possibly one with zero Riemann curvature to one with non-zero.
Thus, eq. \eqref{eq1} should be understood as $Diff(\mathcal{M}) \ni \textbf{d}: \mathcal{F}(g,\phi_b) \to \mathcal{F}(g,\phi_r)$.\footnote{To be rigorous, equation \eqref{eq1} should be understood as $Diff(\mathbb{R}^4) \ni \textbf{m}^{-1}: \mathcal{F}(g,\phi_b) \to \mathcal{F}(g,\phi_r)$. But we have seen that the action of the map \textbf{m} can be written in terms of the action of the external diffeomorphism \textbf{d}, which is its active counterpart. The choice of the term \lq{}\textit{external}\rq{} is related to its use in the literature on subsystems. In particular, we can imagine each set of scalar fields and the metric as dynamically coupled subsystems.  For more on the link between reference frames and subsystems, please refer e.g. to \cite{Belot2017,Wallace2022b,Wallace2022c,GomesSamediff1}.}  So, the upshot is: $\mathcal{F}(g,\phi_b)$ is $d$-invariant, but \textbf{d}-covariant.\footnote{See footnote \ref{fninvariance}. The existence of the external diffeomrphism \textbf{d} proves that choosing a reference frame does not break the covariance of the theory.} In the following, Figure \ref{fig1}  proposes a graphical view akin to Figure \ref{figdvsd}, which compared to Figure \ref{figdvsd} makes it easier to understand the difference between \textbf{d} and $d$.

\subsection{The case of \textbf{URFs}}\label{sec2.2}

What can we say about \textbf{URFs}? Let's consider again the case of a single GPS reference frame $\{\phi^{(I)}\}$.
In that case, it is still true that, given the same initial value problem, one has an entire class of diff-related metrics which can in principle be chosen in order to write different \textbf{(RI)}, but not \textbf{(DET)}, local quantities $\{g_{IJ}(\phi), [d^*g]_{IJ}(\phi), \cdots\}$.\footnote{Here, for notational compactness,  with $[d^*g]_{IJ}(\phi)$ we mean the action of $d$ \textit{only} on the metric: $\big[\phi^{-1}\big]^*[d^*g]_{ab}$.} Here, these observables represent different solutions, and yet their uncoupled nature ensures that they can be obtained from each other by a suitable action of a diffeomorphism.
But since \textbf{(DET)} is not fulfilled, the requirement of \textbf{(GI)} is not met and $\{g_{IJ}(\phi),[d^*g]_{IJ}(\phi),\dots\}$ are not \textit{bona-fide} complete observables.

One might ask whether the absence of deterministic evolution is a real threat to the theory, if it is, so to speak, a pernicious form of indeterminism, or whether it only is an \lq{}apparent\rq{} indeterminism: a discussion that mirrors that surrounding the \textbf{SHA}.

We now introduce a \lq{}\textit{New Hole Argument}\rq{} (\textbf{NHA}) dilemma:
\begin{description}
    \item[\textbf{NHA:}] Given some initial data $\Delta_g$ for the metric, is the indeterminism in the evolution of $g_{IJ}(\phi)$ (for $\phi$ an \textbf{URF}) physically pernicious? Or, put differently, do the two local \textbf{(RI)} representations of the metric $g_{IJ}(\phi),[d^*g]_{IJ}(\phi)$,\textit{ which share the same initial data}, represent two different physical state of affairs? 
\end{description}

We have two possible answers, depending on what one chooses to call \lq{}physical\rq{}:\footnote{Notice that for either \textbf{(i)} or \textbf{(ii)} below, also in the case of \textbf{URFs} there is no underdetermination problem left open, since as shown above in the case of \textbf{CRFs} it is sufficient to have \textbf{(RI)} quantities to solve this problem.}
\begin{description}

\item[\textbf{i) Physical means (GI):}]
In this case, it turns out that the redundancy in the choice of metrics is not a relevant concern, since it does not qualify as a physical indeterminism and there is no \textbf{NHA} left.
Simply, the idea behind this conclusion is that two solutions related by a diffeomorphism acting on only the metric or only on the GPS frame fields do not represent two different physical possibilities since the frame representations of the metric are not \textbf{(GI)}, and so there is no \textit{physical} indeterminism in the dynamics. The situation is completely analogous to the standard Leibniz Equivalence response to the \textbf{SHA}, in which there are two diffeomorphism-related solutions $g_{ab}(p)$ and $[d^*g]_{ab}(p)$ and they can be conceived of as two different mathematical representations of the same physical possibility (\cite{Earman1987-EARWPS}).\footnote{There are other representational possibilities for models in GR. See \cite{Fletcher2020-FLEORC}.} 

%our distinction between \textbf{(GI)} and \textbf{(RI)} allows to interpret a physical situation as only representable by \textbf{(GI)} quantities, i.e. respecting both the demands (\textbf{RI}) \textit{and} (\textbf{DET}). It is not enough to be only \textbf{(RI)}. Consequently, it is straightforwardly true that $(g_{AB}(\phi), [d^*g]_{AB}(\phi))$ represent the \textit{same} physical situation. It is worth noting that, in this case it \textit{does} makes sense to say that they represent the \textit{same} fully-fledged physical situation, since their being \textbf{(RI)} is not sufficient to define them as truly physical.
Consequently, the failure of \textbf{(DET)}, even if accompanied by \textbf{(RI)}, can be understood as bestowing a \lq{}second tier\rq{} unphysical status to the differences between $g_{IJ}(\phi)$ and $[d^*g]_{IJ}(\phi)$. So there is no fully-fledged, `first tier' physical indeterminism.

\item[\textbf{ii) Physical means (RI):}]\footnote{Notice that, if physical means \textbf{(GI)}, than it also means \textbf{(RI)}. The converse is not true.}
In that case the indeterminsm is physically pernicious. We are opening a \textbf{NHA}. The supporter of this stance notes that by being instantiated, \textbf{URFs}, unlike coordinates (or manifold points), give the quantity $g_{IJ}(\phi)$ a physical, relational status that $g_{ab}(p)$ lacks.  Nonetheless, from the same initial data we have an infinite number of diffeomorphism-related solutions, each representing a \textit{bona-fide} physical state of affairs.
Thus, this stance is quite analogous to the haecceitist's endorsement of the \textbf{SHA}.
At the root of that endorsement is the physical meaning  that the haecceitist assigns to points of $\mathcal{M}$. Analogously, here the problem of \textit{physical} indeterminism arises from the fact that \textbf{URFs}, like points of $\mathcal{M}$, are \textit{dynamically} uncoupled from the metric. But unlike points of $\mathcal{M}$ , \textbf{URFs} are instantiated and thus define positions relationally. Assuming that $g_{IJ}(\phi)$ directly represents a state of affairs is in this respect equivalent to assuming that $g_{ab}(p)$ directly represents a state of affairs, and interpreting both $g_{IJ}(\phi)$ and $[d^*g]_{IJ}(\phi)$ as \lq{}physical\rq{} amounts to introducing physical indeterminism into the theory.
%In fact, here the dynamical \textbf{(RI)} quantities represent actual physical possibilities; which is not the case for quantities like $g_{ab}(p)$.
%In general, it is true that what to consider \lq{}physical\rq{} can be ambiguous, but here we point out that when using \textbf{URFs}, if one considers it sufficient to be a \textbf{(RI)} quantity to be a complete observable, then one has to deal with an indeterminism from which there is no escape.
The above is a further argument in support of the need to separate relationalism and gauge-invariance, in support of \cite{BamontiGomes2024} . In particular, choosing one or the other notion as a synonym for \lq{}physical\rq{} has relevant consequences for a theory.

As hinted above, this option of considering a \textbf{(RI)} quantity as physical,  provides an interesting way to analyse the metaphysical concept of \textit{haecceitism}. In particular, it suggests a possible marriage between the haecceitist and the relationalist about spacetime.
For if two \textbf{(RI)} quantities $g_{IJ}(\phi)$ and $[d^*g]_{IJ}(\phi)\equiv g_{IJ}(d^*\phi)$ represent two distinct physical possibilities, then it is possible to use $\phi$ to distinguish a region $U \subset \mathcal{M}$ of the manifold identified by $\phi(U)$ from that by $d^*\phi(U)$.
%{\old However, this physical distinction need not imply  direct  observability (for instance, under the assumption that observability is a dynamical process, i.e. the \lq{}unobservability thesis' of \cite{Wallace2022})\footnote{\old The Unobservability Thesis goes as follows: \lq{}\lq{}Given a family of models of a system which are related by a symmetry transformation, it is impossible to determine empirically which model in fact represents the system \rq{}\rq{} (ivi, page 329). In practice, it is a request that empirical data be relational; a request that has been used several times in this work.} or determinism}. 
At the same time, we are still adopting a relational localisation, where \lq{}this physical point\rq{} acquires a relational meaning as a coincidence of values between the fields $g_{ab}$ and $\phi^{(I)}$.

\end{description}

\begin{comment}
To provide a more comprehensive understanding, let us delve into why, specifically for \textbf{CRFs} (refer to Section \ref{sec2.1}) and case \textbf{(ii)} for \textbf{URFs}, the presence of a coordinate transformation linking two frame-representations of the metrics does not necessarily mean that the sets $\{g_{IJ}(\phi_r),g_{IJ}(\phi_b)\}$ and $\{g_{IJ}(\phi),[d^*g]_{IJ}(\phi)\}$ are physically equivalent. This is in contrast to case \textbf{(i)} applicable to \textbf{URFs}, where such a transformation does imply their physical equivalence.
The point is that in GR, under the understanding of \lq{}physical\rq{} as \textbf{(GI)}, the existence of a coordinate transformation (more generally of a diffeomorphism) relating two \textbf{(GI)} quantities is not a sufficient condition for those quantities to represent the same physical state of affairs.
The \textit{sufficient (and necessary)} condition for two \textbf{(GI)} quantities to be physically the same is the sharing of \textit{the same initial data} set plus the postulate that classical physics is deterministic.
\end{comment}

\begin{figure}[h!]
    \centering
    \includegraphics[scale=0.35]{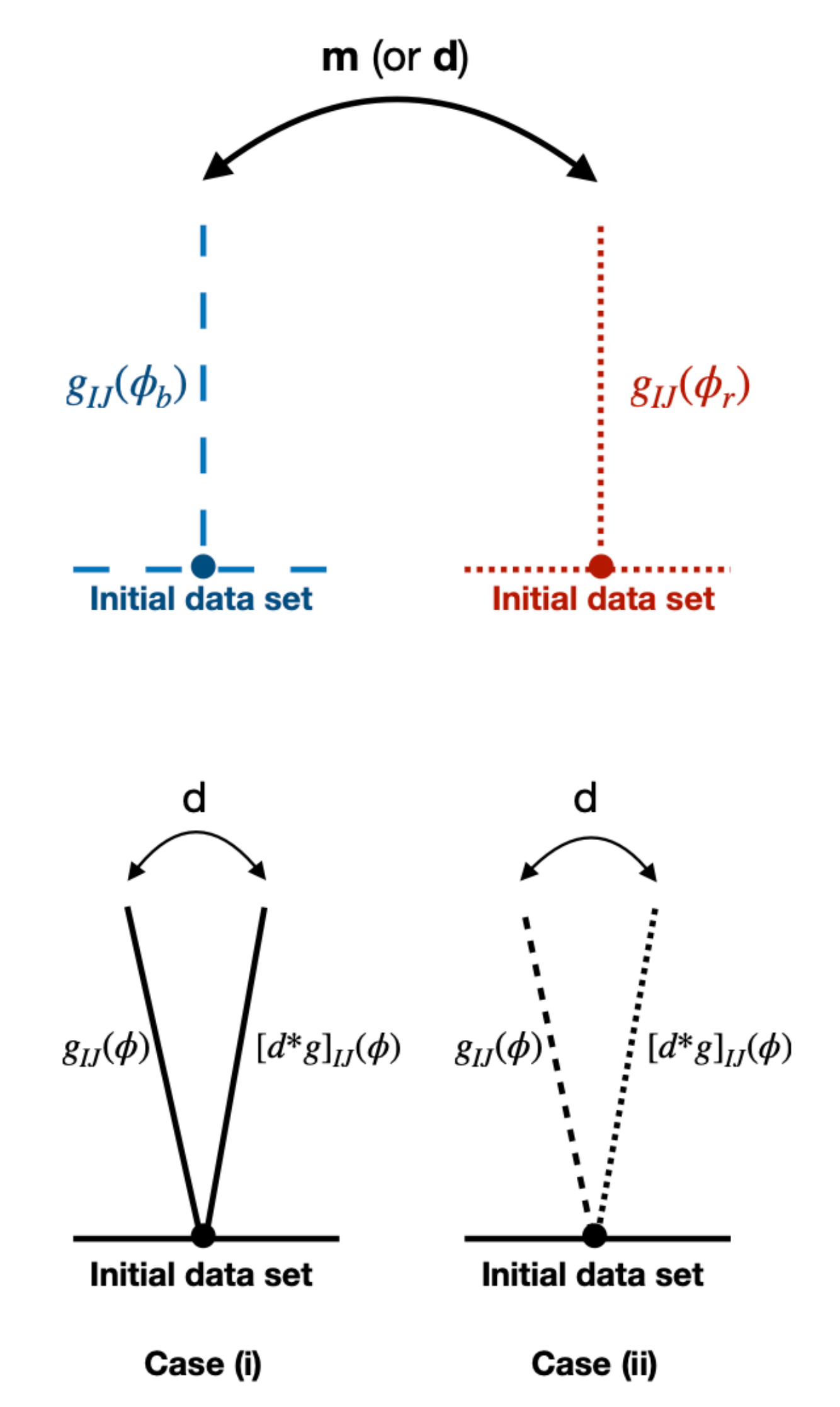}
    \caption{In the figure at the top, we show the case of GPS observables of section \ref{sec2.1}. Starting from two distinct initial data sets (horizontal lines), we have two distinct observables: the blue one (dashed vertical line) and the red one (dotted vertical line), which represent two distinct physical situations. The map \textbf{m} is the passive diffeomorphism of equation \eqref{eq1}. In the figure at the bottom, we have the two cases involving the use of \textbf{URFs}. In case \textbf{(i)} the two quantities represent the same physical situation. In case \textbf{(ii)} the two quantities represent two distinct physical situations: one represented by the dashed line and one by the dotted line. Since the initial data are the same, we have indeterminism.}
    \label{fig3}
\end{figure}

For the sake of clarity, we graphically summarise in Figure \ref{fig3} below the situations set out in this Section, concerning the use of \textbf{CRFs} and \textbf{URFs} (cases \textbf{(i)} and \textbf{(ii)}).

We conclude our analysis noticing that the use of \textbf{URFs} provides also a counterexample of \cite[p.327-328]{Wallace2022} \lq{}Unobservability Thesis\rq{}: \lq{}\lq{}Given a family of models of a system which are related by a symmetry transformation, it is impossible to determine empirically which model in fact represents the system\rq{}\rq{}. This thesis is based on the assumption that empirical access, the process of measurement, is itself a dynamical notion. Thus, the values of (dynamical) symmetry-\textit{variant} quantities cannot be inferred from a dynamical process (viz. a measurement procedure). While this is true for quantities written in some coordinates, like $g_{\mu\nu}(x^\mu)$, it is no longer true for quantities written in \textbf{URFs}, like $g_{IJ}(\phi)$. In fact, although such quantities are (dynamical) symmetry-variant, empirical data \textit{can} distinguish between $g_{IJ}(\phi)$ and $g_{IJ}(d^*\phi)$, since they are \textbf{(RI)} quantities (i.e. quantities invariant under spacetime symmetries \cite{Earman1992-jn}).

\section{Summary}\label{conclusion}

The main results of this work are as follows:

We have addressed the well-known and longstanding conundrum of the Hole Argument in the light of \textbf{CRFs} and \textbf{URFs}. Specifically, we have examined, following the analysis of \cite{HOLE}, both the issues of indeterminism and underdetermination.

In the case of \textbf{CRFs}, using two distinct sets of GPS reference frames as a case study, it becomes immediately apparent that both issues are resolved. However, we have identified a further issue, \textbf{ARB}, consisting of the arbitrary choice between the two sets of frames. Yet, since two complete observables are linked by a diffeomorphism \textbf{d} (also called \textit{external diffeomorphism)}, \textbf{ARB} poses no real threat. We have a \lq{}dictionary\rq{} to translate one observable into the other; the arbitrariness of the choice is not problematic. (Section \ref{sec2.1}).

In the case of \textbf{URFs} we have shown that the hole argument does not lead to a pernicious indeterminism if one interprets as physical only those quantities which are \textbf{(GI)}.
However, if physical status is guaranteed by \textbf{(RI)}, we have a pernicious indeterminism in the theory and the presence of a new hole argument (\textbf{NHA}). On the upside, this position gives a relational instantiation of haecceitist ideas about spacetime. Finally, as in the case of \textbf{CRFs}, since the quantities under consideration are \textbf{(RI)}, there is no underdetermination problem left unresolved. (Section \ref{sec2.2}).

\clearpage
\bibliography{BIB2.bib}
\end{document}